\documentclass{article}
\usepackage{arxiv}
\usepackage[super,sort&compress,comma]{natbib}
\usepackage[version=3]{mhchem}
\usepackage{balance}
\usepackage{verbatim}
\usepackage{times,mathptmx}
\usepackage{sectsty}
\usepackage{graphicx}
\usepackage{lastpage}

\usepackage{float}
\usepackage{fancyhdr}
\usepackage{fnpos}
\usepackage[english]{babel}
\usepackage{soul}
\usepackage[normalem]{ulem}
\usepackage{array}
\usepackage{droidsans}
\usepackage{charter}
\usepackage[T1]{fontenc}
\usepackage[usenames,dvipsnames]{xcolor}
\usepackage{setspace}
\usepackage[compact]{titlesec}
\usepackage{caption}
\definecolor{cream}{RGB}{222,217,201}

\usepackage[utf8]{inputenc}
\usepackage{booktabs}
\usepackage{amsfonts}
\usepackage{amsmath}
\usepackage{xcolor}
\usepackage{hyperref}
\hypersetup{
    colorlinks=true,
    linkcolor=black,
    citecolor=black,
    urlcolor=blue,
}

\newcommand{\eg}{\textit{e.g.}}
\newcommand{\ie}{\textit{i.e.}}

\newcommand{\ModelName}{\textsc{Deep Supervised Graph Partitioning Model}}
\newcommand{\MN}{\textit{DSGPM}}
\newcommand{\MNNOIT}{DSGPM}
\newcommand{\DatasetName}{Human-annotated Mappings}
\newcommand{\DN}{\textit{HAM}}
\newcommand{\DBN}{1,206}

\title{Graph Neural Network Based Coarse-Grained Mapping Prediction}

 \author{
     Zhiheng Li\thanks{These authors contributed equally to this work}\\
     Department of Computer Science\\
     University of Rochester
     \And
     Geemi P. Wellawatte\(^*\)\\
     Department of Chemistry\\
     University of Rochester
     \And
     Maghesree Chakraborty\\
     Department of Chemical Engineering\\
     University of Rochester
     \And
     Heta A. Gandhi\\
     Department of Chemical Engineering\\
     University of Rochester
     \And
     Chenliang Xu\thanks{chenliang.xu@rochester.edu}\\
     Department of Computer Science\\
     University of Rochester
     \And
     Andrew D. White\thanks{andrew.white@rochester.edu}\\
     Department of Chemical Engineering\\
     University of  Rochester\\
 }

\date{}

\begin{document}
\maketitle

\begin{abstract}
    The selection of coarse-grained (CG) mapping operators is a critical step for CG molecular dynamics (MD) simulation. It is still an open question about what is optimal for this choice and there is a need for theory. The current state-of-the art method is mapping operators  manually selected by experts. In this work, we demonstrate  an automated approach by viewing this problem as supervised learning where we seek to reproduce the mapping operators produced by experts. We present a graph neural network based CG mapping predictor called \ModelName (\MN) that treats mapping operators as a graph segmentation problem. \MN{} is trained on a novel dataset, \DatasetName{} (\DN), consisting of \DBN{} molecules with expert annotated mapping operators. \DN{} can be used to facilitate further research in this area. Our model uses a novel metric learning objective to produce high-quality atomic features that are used in spectral clustering. The results show that the \MN{} outperforms state-of-the-art methods in the field of graph segmentation.  Finally, we find that predicted CG mapping operators indeed result in good CG MD models when used in simulation.
\end{abstract}

\newpage

\section{Introduction}

Coarse grained (CG) models can be viewed as a two part problem of selecting a suitable CG mapping and a CG force field. In this work we focus on addressing the issue of CG mapping selection for a given system. A CG mapping is a representation of how atoms in a molecule are grouped to create CG beads. Once the CG mapping is selected, CG force field parameters required for the CG simulation can be determined via existing bottom-up\cite{voth2005} or top-down\cite{xavier} CG methods. The former use atomistic simulations for parameterization of the CG force fields while the latter use experimental data.

Conventionally, a CG mapping for a molecule is selected using chemical and physical intuition. For example, the widely used MARTINI CG model uses mapping of four heavy (non-hydrogen) atoms to one CG bead as chosen by experts\cite{Marrink2013}. Another popular choice of CG mapping for proteins and peptides is to assign one CG bead centered at the \(\alpha\)-carbon for each amino acid. These choices are not built on any thermodynamic or theoretical argument. A recent discussion on commonly used mapping strategies is summarized in \citet{Ingolfsson2014}. There have been recent efforts in developing systematic and automated methods in selecting a CG mapping for a molecule. Automation of CG mapping is important to enhance scalability and transferability.

\citet{Webb2019} proposed a spectral and progressive clustering on molecular graphs to identify vertex groups for subsequent iterative bond contractions that lead to CG mappings with hierarchical resolution. \citet{Wang2019} developed an auto-encoder based method that simultaneously learns the optimal CG mapping of a given resolution and the corresponding CG potentials. \citet{mappingEnt} proposed a mapping entropy based method to simplify the model representation of biomolecules. Their thoeretical model focuses on preserving most information content in the lower resolution model compared to the all atom model.  \citet{Chakraborty2018b} reported a hierarchical graph method where multiple mappings of a given molecule are encoded in a hierarchical graph, which can further be used to auto-select a particular mapping using algorithms like uniform-entropy flattening\cite{Xu2013}. In a recent systematic study on the effects of CG resolution on reproducing on and off target properties of a system, \citet{Khot2019} hypothesized that low-resolution CG models might be information limited, instead of having a representability limitation. This hypothesis suggests that there might be ways of enhancing the information of CG models without increasing their dimension and complexity. This is supported by a recent study of 26 CG mappings for 7 alkane molecules that found little correlation between mapping resolution and CG model performance \cite{Chakraborty2020b}. There is not only a lack of methods to compute mapping operators, there is no agreed upon goal in choosing mapping operators.

Mapping operators used in practice for CG simulations are usually rule-based~\cite{Marrink2013,Ingolfsson2014},but recent advances have been made in algorithmic~\cite{Chakraborty2018b, Webb2019,Zhang2008,Arkhipov2006,Gohlke2006,Li2016} and unsupervised methods\cite{Gomez-Bombarelli2018}. Rule-based schemes have fixed resolution and must be created for each molecular functional group, limiting their application to sequence-defined biomolecules or polymers. Algorithmic and unsupervised methods have only been qualitatively evaluated on specific systems. The \citet{Chakraborty2018b, Gomez-Bombarelli2018} methods also required explicit molecular dynamics simulations, which leads to questions about hyperparameters (e.g., sampling, atomistic force field) and requires at least hours per system. Such methods also are not learning nor optimizing mapping operator correctness directly. Supervised learning has not been used in previous work because there are no datasets and no obvious optimality criteria.

Here we have avoided the open question of "which is the best mapping?", by choosing to match human intuition, the main  selection method of past mapping operators. We demonstrate a supervised learning based approach using a graph neural network framework, \ModelName{}(\MN). To train and evaluate the \MN, we compiled a  \DatasetName{} (\DN{}) dataset with expert annotated CG mappings of \DBN{} organic molecules, where each molecule has one or more coarse graining annotations by human experts. We expect this dataset can facilitate research on coarse graining and the graph partition problem. The \DN{} database allows \MN{} to learn CG mappings directly from annotations. Our framework is closely related to the problem of graph partitioning and has molecular feature extraction and embedding as major components. The graph neural network is trained via metric learning objectives to produce good atom embeddings of molecular graph, which creates better affinity matrix for spectral clustering~\cite{Ng_2002_NeurIPS}. Should there be a consensus in the field on what are ``best'' mappings, our model can be easily adapted to match a new dataset of annotations.

\section{Related Work}

\begin{figure*}[h]
    \centering
    \includegraphics[width=0.85\linewidth]{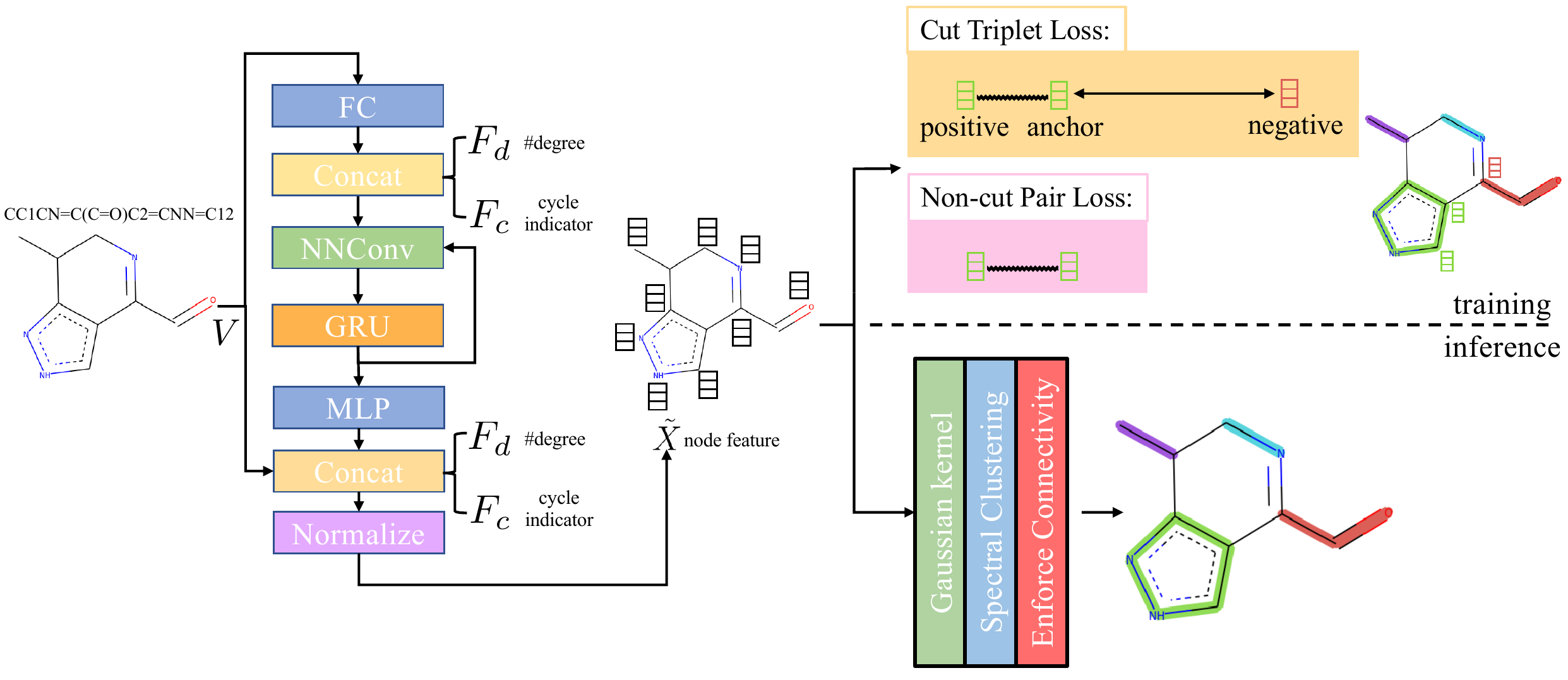}
    \caption{Overview of the method. Adjacency matrix \(E\) is omitted from the figure. FC stands for fully-connected layer and MLP stands for multilayer perceptron. Concat denotes concatenation. NNConv and GRU are explained in Eq.~\ref{eq:nnconv} and Eq.~\ref{eq:gru}, respectively. ``Normalize'' means \(L_2\) normalization.}
    \label{fig:method}
\end{figure*}

\subsection{Molecular Feature Extraction}

The applications of graph convolutional neural networks (GCNN) to molecular modeling is an emerging approach for ``featurizing'' molecular structures. Featurizing a molecule is a challenging process which extracts useful information from a molecule to a fixed representation. This is important since conventional machine learning algorithms can accept only a fixed length input. However, a molecule can have arbitrary sizes and varying connectivities. GCNNs have become a useful tool for molecular featurization as they can be used for deep learning of raw representations of data which are less application specific unlike molecular fingerprints. \citet{Kearnes2016} have shown in their work that GCNNs can be used to extract molecular features with little preprocessing as possible. Furthermore, it is shown that the results from the GCNN are comparable to neural networks trained on molecular fingerprint representations. \citet{moleculenet} have implemented a GCNN featurization method in MoleculeNet. The GCNN is used to compute an initial feature vector which describe an atom's chemical environment and a neighbor list for each atom\cite{moleculenet}. Additionally, they show that unlike the fingerprints methods, GCNNs create a learnable process to extract molecular features using differentiable network layers. \citet{Gilmer2017} have developed a generalized message passing nerural network (MPNN) to predict molecular properties. In this work, authors have used a GCNN to extract molecular features and to learn them from molecular graphs. The authors also state that there is a lack of a generalized framework which can work on molecular graphs for feature extraction. Given the proven success of GCNNs in feature extraction, the motivation for our work was to develop a generalized deep learning based method apt for chemistry problems.

\subsection{Graph Partitioning and Graph Neural Network}

If a molecule is viewed as a graph, the problem of selecting a CG mapping is analogous to partitioning the molecular graph. While there has been limited application of molecular graph for the purpose of selecting CG mappings, as discussed earlier, we would like to highlight some strategies employed for problems relevant to graph partitioning. Spectral clustering~\cite{Weiss_1999_ICCV,Shi_PAMI_2000,Ng_2002_NeurIPS} is one of the baseline method used in graph clustering task. Compared with Expectation-Maximization (EM)~\cite{Dempster_1977_JRSS}, spectral clustering has a better modeling on pairwise affinity given by the adjacency matrix of a graph. METIS~\cite{Karypis_1998_SIAM} solves the graph partition problem in a multilevel scheme via coarsening, partition, and refinement steps. Graclus~\cite{Dhillon2007} proposed a generalized kernel k-means method with better speed, memory efficiency, and graph clustering result. \citet{Fortunato2010} has a comprehensive review of the methods developed for community detection in graphs. \citet{Safro2012} compared different graph coarsening schemes for graph partitioning using algebraic distance between nodes of the graph. Recently, some graph neural network~\cite{Battaglia_2018_ArXiv} based graph partitioning methods have been proposed. GAP~\cite{Nazi2019} uses graph neural networks to predict node-partition assignment probability, which is learned through normalized cut loss and balanced cut loss. ClusterNet~\cite{Wilder_2019_NeurIPS} adds differentiable k-means clustering at the end of graph neural network to enable end-to-end training. Compared to the aforementioned methods,  our \MN{} combines the advantages of both spectral clustering and a graph neural network, leading to better results than either alone. We also propose and justify a novel metric learning loss to train the graph neural network.

\subsection{Metric Learning}
The goal of the metric learning is to learn a model which encodes the input data to an embedding space, where embeddings (usually represented by fixed length vector) of similar data objects are separated by short distances in the embedding space and different data objects are separated by larger distances in the embedding space. \citet{Hadsell_2006_CVPR} proposed a siamese network trained via contrastive loss which 1) minimizes \(L_2\) distance for instances from the same group, and 2) maximizes \(L_2\) distance for instances from the different groups if the \(L_2\) distance is larger than a margin. Instead of only considering a pair of data, \citet{Schroff_2015_CVPR} considered a triplet of data \(\langle\)anchor, positive, negative\(\rangle\) and triplet loss to ensure \(L(\text{anchor}, \text{negative})\) (distance between anchor and negative) should be larger than \(L(\text{anchor}, \text{positive})\) (distance between anchor and positive) by a margin. However, the methods above have only been applied to nonstructural data (\eg, image clustering). Furthermore, one of challenging problem is sampling pairs or triplets of data from the dataset. In contrast, our proposed method can efficiently enumerate pairs or triplets by explicitly treating the graph structure.

\section{Method}
\subsection{Problem Formulation}
\ModelName{} (\MN)~\footnote{ The code for \MN{} can be accessed via \url{https://github.com/rochesterxugroup/DSGPM}.} formulates the CG mapping prediction as a graph partitioning problem. Suppose \(Q\) is the set of atom types existing in the dataset. An atom in a molecule is represented as a one-hot encoding of its atom type. Similarly, a bond is represented as a one-hot encoding of its bond type (\eg, single, double, aromatic, etc.). Therefore, a molecule with \(n\) atoms is formulated as a graph \(G = (V, E)\) , where \( V \in \mathbb{R}^{n \times |Q|} \) represents atoms and \(E \in \mathbb{R}^{n \times n \times 4}\) denotes the adjacency matrix with encoded bond types.

\subsection{Motivation}
One strong baseline method to solve the graph partitioning problem is spectral clustering~\cite{Ng_2002_NeurIPS}. The performance of spectral clustering is mainly decided by the quality of affinity matrix \(S \in \mathbb{R}^{n \times n}\), where \(S_{ij}\) denotes the affinity (ranging from 0 to 1) between vertex \(i\) and vertex \(j\). In this task, the adjacency matrix (ignoring bond type information in \(E\)) can serve as the affinity matrix fed into spectral clustering. However, for the CG mapping prediction problem, an ideal affinity matrix should have low affinity value of cut (edge connecting two atoms from different CG beads) and high affinity of an edge which is not a cut, while adjacency matrix only contains ``0''s and ``1''s to represent the existence of edges.

\subsection{\ModelName{}}
The main difference from the baseline method is a graph neural network \(\mathcal{F}\) that is used to obtain a better affinity matrix, where \(\mathcal{F}\) follows the architecture design of MPNN~\cite{Gilmer2017}. The overview of the method is shown in Fig.~\ref{fig:method}. With the molecular graph \(G\) as the input, \(\mathcal{F}\) extracts \(q\)-dimensional atom features \(\tilde{X} \in \mathbb{R}^{n \times q}\) through message passing (\ie, \(\tilde{X} = \mathcal{F}(G)\)). Concretely, \(\mathcal{F}\) first uses a fully-connected layer to project one-hot atom type encoding into the feature space. Then, we concatenate the embedded feature with two numbers: 1) number of degree; 2) cycle indicator (\ie, whether the atom is in a cycle) (zero or one) to obtain \(d\)-dimensional feature \(X^0 \in \mathbb{R}^{n \times d} \). We find out that adding these two features improves the result (Sec.~\ref{subsec:ablation}). Next, \(X^0\) is iteratively updated \(T\) time steps to obtain \(X^T\):

\begin{align}
    \label{eq:nnconv}
    \hat{X}^{t-1}_u &= \mathbf{W}' X^{t-1}_u + \sum_{v \in \mathcal{N}(u)} X^{t-1}_v \phi^e(E_{uv}), \\
    X^t_u &= \text{GRU}(\hat{X}^{t-1}_u, H^{t-1}_u),
    \label{eq:gru}
\end{align}
where underscript \(u\) denotes \(u\)-th atom and superscript \(t\) denotes time step; \(\mathcal{N(\cdot)}\) denotes the set of neighboring nodes of the given vertex; \(\mathbf{W}  \in \mathbb{R}^{d \times d}\) is a weight matrix; superscript \('\) denotes transpose; \(\phi^e(\cdot): \{0, 1\}^{4} \mapsto \mathbb{R}^{d \times d}\) is function mapping bond type to edge-conditioned weight matrix, which is implemented as multilayer perceptron; GRU stands for Gated Recurrent Unit~\cite{Cho_ACL_2014}. Finally, the output feature \(\tilde{X}\) is obtained by:

\begin{align}
    \tilde{X}' &= \text{Concat}(\text{MLP}(X^T), V, F_d, F_c), \\
    \tilde{X} &= \frac{\tilde{X}'}{||\tilde{X}'||_2},
\end{align}
where Concat denotes concatenation; MLP denotes multilayer perceptron; \(F_d \in \mathbb{N}^n\) denotes degree of each atom and \(F_c \in \{0, 1\}^n\) is cycle indicator (\ie whether an atom is in a cycle).

After computing the atom feature \(\tilde{X}\), the affinity matrix \(A \in \mathbb{R}^{n \times n}\) can be calculated by a Gaussian kernel:
\begin{equation}
    A_{ij} = \exp \left(\frac{-||\tilde{X}_i - \tilde{X}_j||_2^2}{2\sigma^2}\right) \tilde{E},
    \label{eq:gaussian}
\end{equation}
where \(\sigma\) is the bandwidth and is set to \(\sigma = 1\) in the experiment. $\tilde{E} \in \mathbb{R}^{n \times n}$ denotes the adjacency matrix ($\tilde{E}_{ij} = 1$ if atom $i$ and atom $j$ are bonded, otherwise $\tilde{E}_{ij} = 0$).

Therefore, in order to obtain a good affinity matrix, \(||\tilde{X}_i - \tilde{X}_j||_2\) should be large when edge \(\langle i, j \rangle\) is a cut and small when edge \(\langle i, j \rangle\) is not a cut. Hence, by utilizing the ground-truth partition \(B \in \mathbb{N}^n\) (\(B_i\) denotes coarse grain (partition) index of \(i\)-th atom), we design \textit{Cut Triplet Loss} and \textit{Non-cut Pair Loss} to guide the network outputting good node feature \(\tilde{X}\) during training.

\subsection{Training}

\subsubsection{Cut Triplet Loss.}
The goal of \textit{Cut Triplet Loss} is to push pairs of node embeddings far away from each other when they belong to different partitions. To this end, we first extract all triplets from the given molecular graph \(G\) where each triplet contains three atoms: (anchor atom, positive atom, negative atom) denoted by \( \{\mathtt{a, p, n}\} \) such that \( B_\mathtt{a} = B_\mathtt{p} \) but \( B_\mathtt{a} \neq B_\mathtt{n} \) (see ``green'' features and ``red'' feature on top-right of Fig.~\ref{fig:method}). In other words, we extract non-cut edge \(\langle \mathtt{a}, \mathtt{p} \rangle\) and cut edge \(\langle \mathtt{a}, \mathtt{n} \rangle\) sharing one common vertex \(\mathtt{a}\). The set of triplets is denoted by \(P\). Then, \textit{Cut Triplet Loss} is defined by:
\begin{equation}
    L_\text{triplet} = \frac{1}{|P|} \sum_{ \substack{P_i = \{\mathtt{a, p, n}\} \\ i \in [1, |P|]}} \text{max}(||\tilde{X}_\mathtt{a} - \tilde{X}_\mathtt{p}||_2 - ||\tilde{X}_\mathtt{a} - \tilde{X}_\mathtt{n}||_2 + \alpha, 0),
\end{equation}
where \(\alpha\) is a hyperparameter denoting the margin in triplet loss. By optimizing \textit{Cut Triplet Loss}, the objective \(||\tilde{X}_\mathtt{a} - \tilde{X}_\mathtt{p}||_2 + \alpha \leq ||\tilde{X}_\mathtt{a} - \tilde{X}_\mathtt{n}||_2\) can be satisfied for all triplets.

\subsubsection{Non-cut Pair Loss.}
The purpose of \textit{Non-cut Pair Loss} is to pull pairs of node embeddings as close as possible when they are from the same partition. Therefore, all pairs of node \(\mathtt{a}\) and \(\mathtt{a}'\) are extracted when edge \(\langle \mathtt{a}, \mathtt{a}' \rangle\) is not a cut. The set of pairs of node is denoted as \(S\). Then, \textit{Non-cut Pair Loss} is defined by:
\begin{equation}
    L_\text{pair} = \frac{1}{|S|} ||\tilde{X}_\mathtt{a} - \tilde{X}_{\mathtt{a}'}||_2 .
\end{equation}

The final loss function to train the network is defined by:
\begin{equation}
    L = L_\text{triplet} + \lambda L_\text{pair},
    \label{eq:final_loss}
\end{equation}
where the coefficient is taken to be $\lambda = 0.1$.

\subsection{Inference}
In the inference stage, we first apply Eq.~\ref{eq:gaussian} on the extracted node feature \(\tilde{X}\). Then, based on affinity matrix \(A\), spectral clustering is used to obtain the graph clustering result. Note that graph clustering is slightly different to graph partitioning. The latter requires the predicted partition must be a connected-component. Hence, we post-process the graph clustering result by enforcing connectivity of each graph partition: for each predicted graph cluster, if it contains more than one connected-component, we assign new indices to each connected-components.

\begin{figure*}[t]
    \centering
    \includegraphics[width=\linewidth]{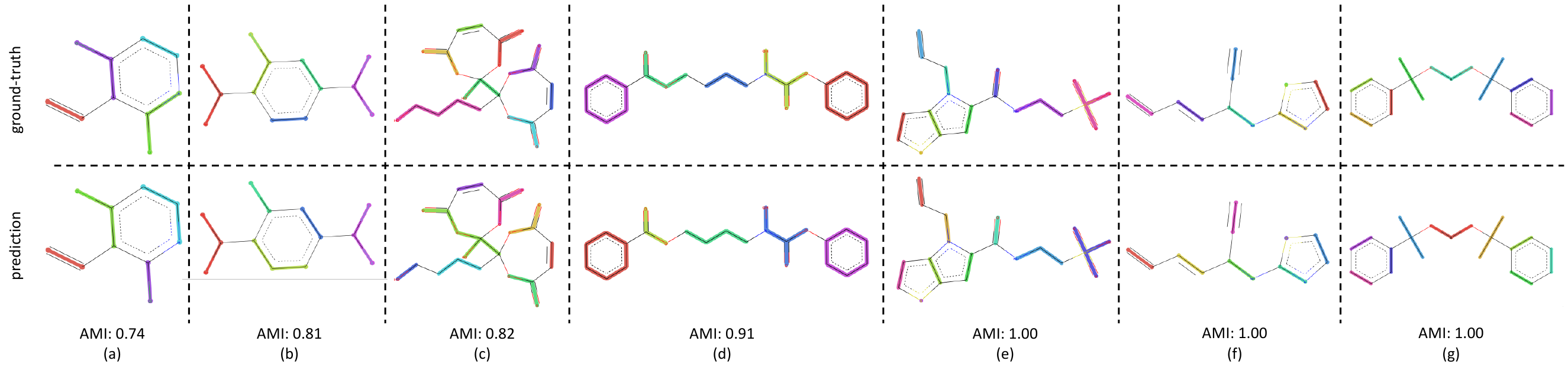}
    \caption{Visualization of the CG mapping prediction and the ground-truth. Atoms and corresponding edges belongs to the same CG bead are highlighted with the same color. Cut edges are not highlighted (\ie, in black). Note that colors between prediction and ground-truth may not match since colors are randomly selected.}
    \label{fig:cg_result}
\end{figure*}

\section{Experiment}
\subsection{Dataset}
\DatasetName{} (\DN) dataset~\footnote{\DN{} dataset can be downloaded via \url{https://github.com/rochesterxugroup/HAM_dataset/releases}} contains CG mappings of \DBN{} organic molecules with less than 25 heavy atoms. Each molecule was downloaded from the PubChem database as SMILES~\cite{pubchem}. One molecule was assigned to two annotators to compare the human agreement between CG mappings. These molecules were hand-mapped using a web-app. The completed annotations were reviewed by a third person, to identify and remove unreasonable mappings which did not agree with the given guidelines. Hence, there are 1.68 annotations per molecule in the current database (16\% removed).
To preserve the chemical and physical information of the all atom structure accurately, the annotators were instructed to group chemically similar atoms together into CG beads while preserving the connectivity of the molecular structure. They were also instructed to preserve the planar configuration of rings if possible by grouping rings into 3 or more beads.

\subsection{Evaluation Metrics}
Adjusted Mutual Information (AMI)~\cite{Vinh_2010_JMLR} is used to evaluate the graph partition result in terms of nodes in the graph. Nodes from the same CG bead are assigned with the same cluster index and AMI compares predicted nodes' cluster indices with ground-truth nodes' cluster indices. We also evaluate graph partition result in terms of accuracy of predicting cuts from a graph. We report the precision, recall, and F1-score on cuts prediction (denoted by Cut Prec., Cut Recall, and Cut F1-score, respectively). Our method is trained and evaluated through 5-fold cross-validation~\cite{kohavi1995study} to mitigate the bias of data split. Concretely, the dataset is split into 5 non-overlapping partitions (\ie, one molecule only exists in one data partition). The experiment will run 5 iterations. At \(i\)-th iteration (\(i \in [1, 5]\)), the \(i\)-th split of the dataset is regarded as testing set (ground-truth partition \(B\) is not used) and rest 4 splits of the dataset is regarded as the training set (ground-truth partition \(B\) is used for training). Therefore, after training, \MN{} is evaluated on \textbf{unseen} molecules in the testing set. The final results are the average values over all iterations. Since one molecule may have multiple annotations, we choose one of the annotations that produces the best result for both AMI and cut accuracy.

\subsection{Implementation Details}
\MN{} is trained with at most 500 epochs and we choose the epoch at which model achieves the best performance over the 5-fold cross validation~\footnote{This setting is also used for comparison methods.}. The hidden feature dimension is 128. The implementation of spectral clustering used in the inference stage is from Scikit-learn~\cite{Pedregosa_2011_JMLR}. Since spectral clustering requires a hyperparameter to indicate the expected number of clusterings, we provide the ground-truth number of clusters based on CG annotations. Cycles of each molecular graph are obtained via ``cycle\_basis''~\cite{Paton_1969_CACM} function implemented by NetworkX~\cite{Hagberg_2008}. The code of graph neural network is based on PyTorch~\cite{Paszke_2019_NeurIPS} and PyTorch Geometric~\cite{Fey_2019_ICLRW}.

\begin{table}[t]
    \centering
    
        \begin{tabular}{l|cccc}
            \toprule
                Method                       & AMI             & Cut Prec. & Cut recall & Cut F1-score \\
            \midrule
                GAP~\cite{Nazi2019}     & 0.33          & 0.47             & 0.73          & 0.54            \\
                Graclus~\cite{Dhillon2007}                     & 0.45           & 0.58             & 0.81          & 0.65            \\
                ClusterNet~\cite{Wilder_2019_NeurIPS}    & 0.52          & 0.64            & 0.62          & 0.58           \\
                METIS~\cite{Karypis_1998_SIAM}                        & 0.56         & 0.63             & 0.56          & 0.58            \\
                Cut Cls.           & 0.67         & 0.75    & 0.73          & 0.73       \\
                Spec. Cluster.~\cite{Ng_2002_NeurIPS}          & 0.73          & 0.75              & 0.75          & 0.75            \\
                \textbf{\MNNOIT} (Ours) & \textbf{0.79} & \textbf{0.80}
                & \textbf{0.80} & \textbf{0.80}   \\ \midrule
                Human              & 0.81 & 0.81    & 0.81 & 0.81 \\
            \bottomrule
            \end{tabular}
    \caption{Comparison with state-of-the-art methods. Average results over 5-fold cross validation are shown. Here, ``Spec. Cluster.'' means spectral clustering. The standard deviation of 5-fold cross-validation result under all evaluation metrics of our method is smaller than 0.01. Evaluation on human agreement (last row) is based on 128 molecules with 129 pairs of mappings, where mappings in each have the same number of CG beads.}
    \label{tab:sota}
\end{table}

\subsection{Comparison with State-of-the-Art}
We compare our method with five state-of-the-art graph partitioning methods. We used officially released code of the comparing methods on \DN{} dataset. Here, we also show an alternative of our method (denoted by \textit{Cut Cls.}): by regarding the graph partitioning problem of edge cut binary classification problem (\ie, predicting the probability that an edge is a cut or not), we train \MN{} with binary cross-entropy loss. In the inference stage, we rank ``cut probability'' of each edge in descending order and take top-\(k\) edges as the final cut prediction, where \(k\) is the ground-truth number of cuts computed from the CG annotation. The result of comparison is shown in Tab.~\ref{tab:sota}. The result shows that our method outperforms all state-of-the-art methods in terms of both AMI and cut accuracy. Moreover, \MN{} also outperforms \textit{Cut Cls.}, proving the effectiveness the metric learning training objectives and the importance of spectral clustering stage in our method. Additionally, by treating one annotation as prediction and the other annotation as ground-truth, we can show the agreement between different annotations (see last row in Tab.~\ref{tab:sota}), which can be regarded as human annotator's performance. The result shows that our proposed \MN{} is very closed to human-level performance.

\begin{table}[t]
    \centering
    
        \begin{tabular}{l|cccc}
            \toprule
            input              & AMI             & Cut Prec.   & Cut Recall      & Cut F1-score    \\ \midrule
            w/o \(F_d\) \& \(F_c\) & 0.781          & 0.797          & 0.801          & 0.798          \\
            w/o \(F_c\)           & 0.783          & 0.800          & 0.803          & 0.801          \\
            w/o \(F_d\)          & 0.790          & 0.806          & 0.807          & 0.806         \\
            \MNNOIT & \textbf{0.790} & \textbf{0.806}
            & \textbf{0.809} & \textbf{0.807}\\ \bottomrule
            \end{tabular}
    \caption{Ablation study on the input of \MN. \(F_d\) and \(F_c\) denote number of degree and cycle indicator, respectively.}
    \label{tab:input_feat_ablation}
\end{table}

\begin{table}[t]
    \centering
    \scalebox{0.9}{
        \begin{tabular}{l|cccc}
            \toprule
            loss terms              & AMI             & Cut Prec.   & Cut Recall      & Cut F1-score    \\ \midrule
            w/o \(L_\text{triplet}\) & 0.73          & 0.75        & 0.76          & 0.7         \\
            w/o \(L_\text{pair}\)           & 0.78          & 0.80          & 0.80          & 0.80          \\
            
            \MNNOIT & \textbf{0.79} & \textbf{0.80}
         & \textbf{0.80} & \textbf{0.80} \\ \bottomrule
            \end{tabular}
    }
    \caption{Ablation study on loss terms. \(L_\text{triplet}\) denotes \textit{Cut Triplet Loss} and \(L_\text{pair}\) denotes \textit{Non-cut Pair Loss}. \label{tab:loss_ablation}}
\end{table}

\begin{table}[t]
    \centering
    \begin{tabular}{l|cccc}
        \toprule
        \(\lambda\)              & AMI             & Cut Prec.   & Cut Recall      & Cut F1-score    \\ \midrule
        0.1 & 0.79         & 0.80        & 0.80          & 0.80           \\
        0.5 & 0.78          & 0.80       & 0.80          & 0.80 \\
        1 & 0.78          & 0.80        & 0.80          & 0.80 \\
        2 & 0.78          & 0.80        & 0.80          & 0.80           \\
        10 & 0.78          & 0.80        & 0.80          & 0.80           \\
     \bottomrule
    \end{tabular}
    \caption{Ablation study on loss terms. \(\lambda\) denotes the coefficient for \textit{Non-cut Pair Loss} (Eq.~\ref{eq:final_loss}). \label{tab:lambda_ablation}}
\end{table}

\begin{table}[t]
    \centering
    \begin{tabular}{l|cccc}
        \toprule
        \(\sigma\)              & AMI             & Cut Prec.   & Cut Recall      & Cut F1-score    \\ \midrule
        0.5 & 0.77          & 0.79        & 0.79          & 0.79 \\
        1          & 0.79 & 0.80        & 0.80          & 0.80           \\
        1.5 & 0.77         & 0.79        & 0.79          & 0.79 \\
         2          & 0.76         & 0.78        & 0.78          & 0.78           \\
     \bottomrule
    \end{tabular}
    \caption{ Ablation study on bandwidth of Gaussian kernel. \(\sigma\) denotes the bandwidth for Gaussian kernel in Eq.~\ref{eq:gaussian}. \label{tab:sigma_ablation}}
\end{table}

\subsection{Ablation Study}
\label{subsec:ablation}

We study the contribution of degree and cycle indicator in the input. The results are shown in Tab.~\ref{tab:input_feat_ablation}. Degree feature (w/o \(F_c\) in Tab.~\ref{tab:input_feat_ablation}) improves the edge-based metrics (cut precision, cut recall, cut F1-score) and cycle indicator (w/o \(F_d\) in Tab.~\ref{tab:input_feat_ablation}) contributes to all evaluation metrics. Combining both input feature boosts the performance further.

We also examined the contribution of each loss terms, cut triplet loss and non-cut pair loss. The result in Tab.~\ref{tab:loss_ablation} shows that \textit{Cut Triplet Loss} plays the major role in the training objective and combining both loss terms will produce better performance, which proves that \(L_\text{triplet}\) and \(L_\text{pair}\) 's objectives, separating atoms connected by an cut edge and concentrating features of atoms from the same partition, are reciprocal during training.

Furthermore, we study the impact of different values for the hyperparameters \(\lambda\) (see Eq.~\ref{eq:final_loss}) and \(\sigma\) (see Eq.~\ref{eq:gaussian}) in Tab.~\ref{tab:lambda_ablation} and Tab.~\ref{tab:sigma_ablation}, respectively. The ablation results show that \MN{} is not sensitive to changes of \(\lambda\) and choosing \(\sigma = 1\) yields best results.

\subsection{Visualization}
\subsubsection{CG Mapping Result}
We visualize the CG mapping prediction results against ground-truth in Fig.~\ref{fig:cg_result}. Predicted mappings \textit{(e)-(g)} are indistinguishable from the human annotations. Even though AMI values of structures \textit{(a)-(c)} are comparatively lower, our predictions in \textit{(a)-(c)} are still able to capture the essential features such as functional groups and ring conformations from the ground truth mappings. \textit{(a),(b),(e)-(g)} also show that when rings in molecules are  grouped into three CG beads by the human annotators, \MN{} model is able to capture this pattern. When rings are grouped into one CG bead (\ref{fig:cg_result} \textit{(d)}), the model similarly chose this. Overall this shows that \MN{} can reproduce mappings which are significantly close to the human annotations.  We have further compared our predictions with the widely used MARTINI mapping scheme. Results are shown in Fig. S3 in the SI.

\subsubsection{SARS-CoV-2 Structure Prediction}
\begin{figure}[htb]
\centering
\includegraphics[width=\linewidth]{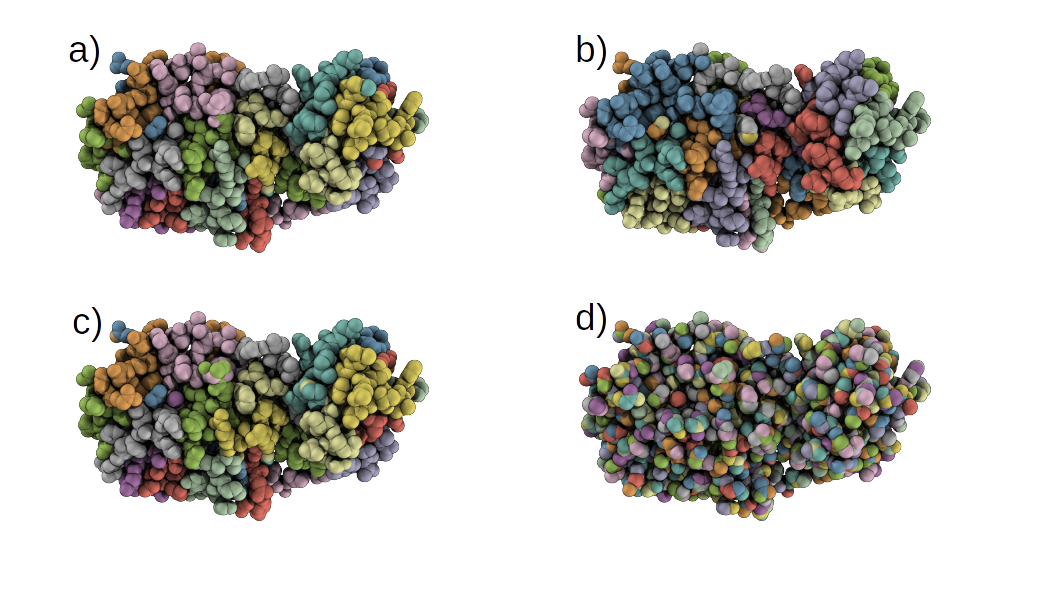}
\caption{Comparison of CG mappings of SARS-CoV-2 protease structure predicted by baseline methods, a) our \MN{} model  b) METIS\cite{Karypis_1998_SIAM}  c) Spectral clustering\cite{Ng_2002_NeurIPS}   d) Graclus\cite{Dhillon2007} . a), b) and c) have 32 CG beads while d) contains 1455 CG beads.}
\label{fig:protein_result}
\end{figure}

Using our trained  \MN{}, we predict the CG mappings for previously unseen SARS-CoV-2 protease structure (PDB ID:6M03\cite{pdb}). In Fig. \ref{fig:protein_result} we compare our result with three baseline methods. Even though our training dataset did not contain peptide sequences we show that our model is capable of predicting CG mappings of complex proteins. We see in Fig. \ref{fig:protein_result} that our prediction is similar to predicted mapping from the spectral clustering method. This is an expected result as we use spectral clustering in the inference stage of our model. In spectral clustering, METIS methods and our model the resolution of the CG mapping can be controlled as the number of partitions is a hyper parameter. Mappings predicted by these three methods in Fig.\ref{fig:protein_result} contain 32 beads. However, in the Graclus method the resolution cannot be controlled. In Fig. \ref{fig:protein_result} d), the predicted mapping from Graclus method contain 1455 CG beads. This is not a reasonable prediction as the fine grain structure contains 2367 atoms.

To gain a better understanding of the mappings, in Fig. S1 in the SI, we use the FASTA representation of the SARS-CoV-2 protease and color each one-letter code by the color of the CG beach to which each alpha-carbon belong. We see that our model is able to group amino acids with reasonable cuts along the backbone of the protein. Our model and spectral clustering method group 7-11 amino acids while the METIS method group 2-11 amino acids into CG beads. This shows that while \MN{} is capable of predicting state-of-the-art mapping for small molecules it can also be scaled to predict reasonable mappings for arbitrarily large structures.

\subsubsection{Model Performance in CG Simulations}
\begin{figure*}[htb]
    \centering
    \includegraphics[width=\linewidth]{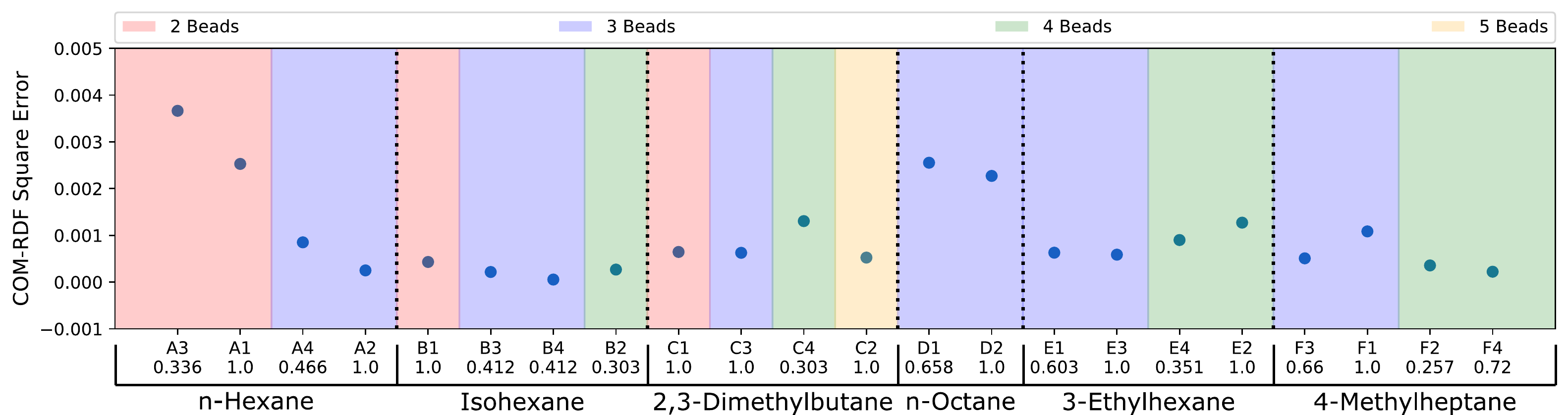}
    \caption{COM-RDF square errors as previously reported for CG mappings of 6 alkane molecules~\cite{Chakraborty2020b}. The mappings for each molecule have been categorized into colored blocks corresponding to the number of CG beads. For each block, the mappings are arranged in the order of increasing AMI values, as indicated below the CG mapping labels.}
    \label{fig:alkane_ami_comparison}
\end{figure*}
Thus far, the model has been judged against \DatasetName{} and not in molecular dynamics simulation. To assess the predicted mappings, we draw upon the simulation results from recent work by \citet{Chakraborty2020b} where force matching was used for coarse-graining. We have compared the performance of the CG mappings predicted by \MN{} for 6 alkane molecules with multiple CG bead numbers, giving 22 different simulation results. The individual mappings of the 6 alkane molecules (n-hexane, isohexane, 2,3-dimethylbutane, n-octane, 3-ethylhexane, and 4-methylheptane) that were considered in \citet{Chakraborty2020b} and those predicted by \MN{} are shown in Fig. S2. \MN{} predicts one mapping per molecule/bead number. To assess the quality of these mappings, we show how the CG simulation error changes for mappings other than the predicted \MN{} mapping as measured by AMI. Decreasing error as a AMI increases (better performance as we get closer to the \MN{} prediction) indicates good model performance. Fig.~\ref{fig:alkane_ami_comparison} shows the square errors for center-of-mass (COM) radial distribution function (RDF) relative to the all atom simulation as previously reported~\cite{Chakraborty2020b} for each of the 6 alkane molecules. For a given molecule, the mappings are categorized into colored blocks corresponding to the number of beads in the CG mapping. AMI values of the mappings are computed relative to the CG mappings from \MN{} with the same number of CG beads. The mappings within the same colored block are arranged in increasing order of AMI values. It is observed that for most of the alkanes, a mapping with higher AMI compared to another with equal number of beads, yields lower COM-RDF square error (6 instances). 4 bead 3-ethylhexane mappings and 3 bead 4-methylheptane mappings are the only instances where a mapping with higher AMI gives higher COM-RDF square error than a comparable mapping with lower AMI. Thus the mappings predicted by \MN{} have good performance when used in simulations as judged from this small dataset of 22 simulations.

\section{Conclusion}
In this work, we propose a novel \MN{} as a supervised learning method for predicting CG mappings. By selecting good inputs and designing novel metric learning objectives on graph, the graph neural network can produce good atom features, resulting in better affinity matrix for spectral clustering. We also report the first large-scale CG dataset with experts' annotations. The result shows that our method outperforms state-of-the-art methods by a predicting mappings which are nearly indistinguishable from human annotations. The ablation study found that the novel loss term is the key innovation of the model. Furthermore, we show that our automated model can be used to predict CG mappings for macromolecules even though the training set was of small molecules and the CG mappings do result in good performance when implemented in force-matched CG simulations.

\section{Conflicts of interest}
There are no conflicts to declare.

\section{Acknowledgements}
This material is based upon work supported by the National Science Foundation under Grant No. 1764415. We thank the Center for Integrated Research Computing at the University of Rochester for providing the computational resources required to complete this study. Part of this research was performed while the authors were visiting the Institute for Pure and Applied Mathematics (IPAM), which is supported by the National Science Foundation (Grant No. DMS-1440415).

\section{Supplementary Information}

\begin{figure}[H]

  \includegraphics[width=\linewidth]{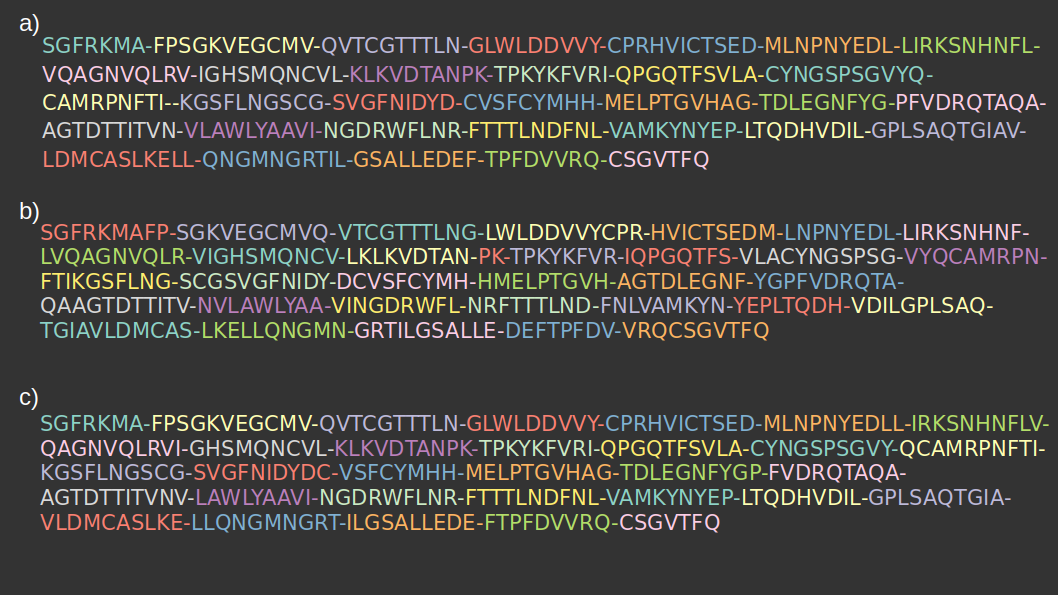}
  \caption{Comparison of FASTA representations of the SARS-CoV-2 main protease coarse grained mappings predicted by  \MN{} model with baseline methods. Predicted mappings from a) our \MN{} model b) METIS\cite{Karypis_1998_SIAM} and c) Spectral clustering\cite{Ng_2002_NeurIPS} are illustrated. All three mappings presented here  have 32 CG beads. We have colored each one-letter label of amino acids by the color of CG bead to which each alpha-carbon belong.}
  \label{fig:seq_result}

\end{figure}

\begin{figure}
\centering
  \includegraphics[width=0.7\linewidth]{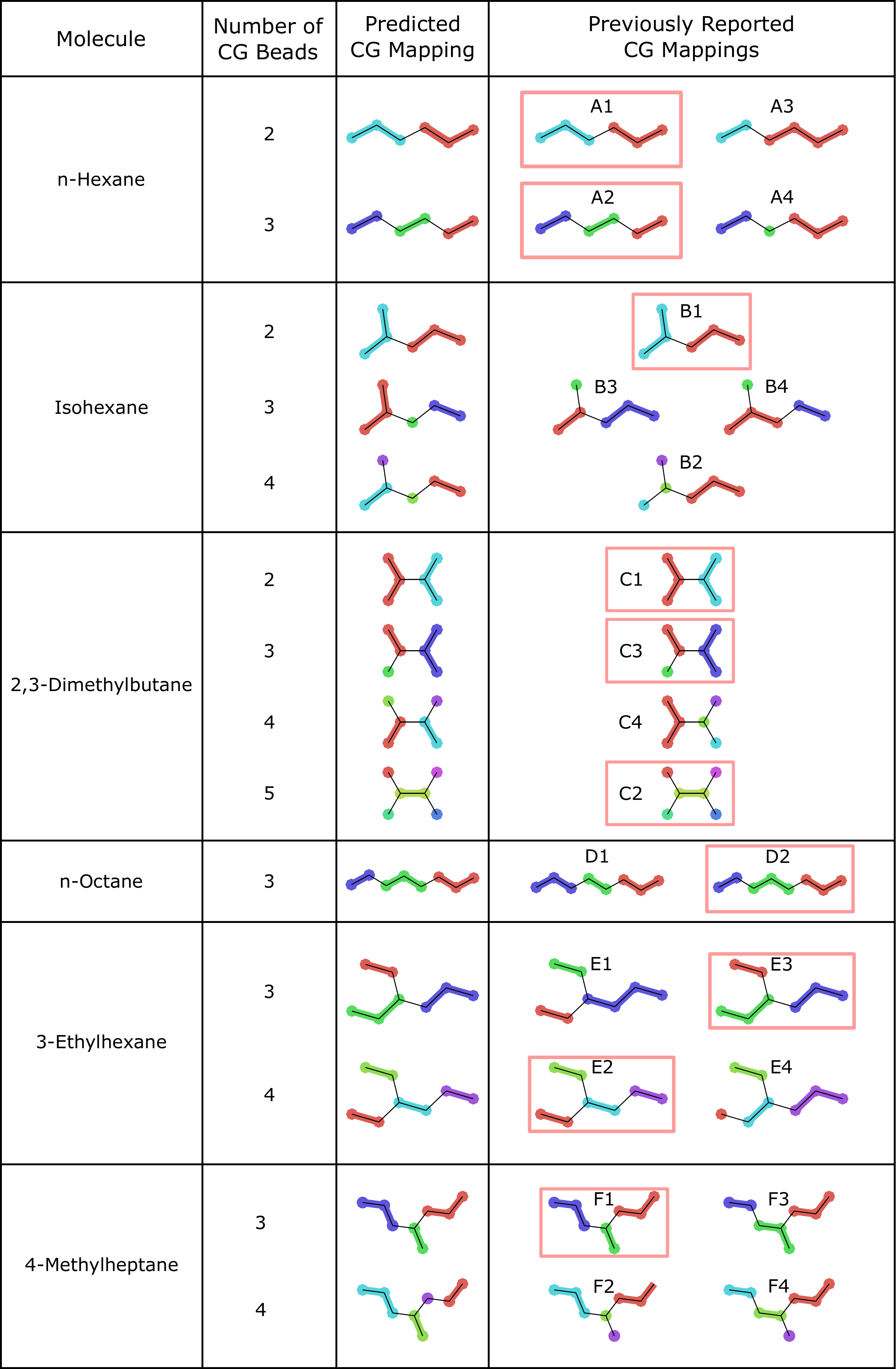}
  \caption{Illustrations of previously studied CG mappings of 6 alkane molecules~\cite{Chakraborty2020b} (n-hexane, isohexane, 2,3-dimethylbutane, n-octane, 3-ethylhexane, and 4-methylheptane) and \MN{} predicted CG mappings of the 6 molecules with varying CG bead number. The mappings enclosed in red boxes correspond highlight instances where the predicted mappings are identical to one of previously studied mappings with the same number of CG beads and hence have AMI value as 1.}
  \label{fig:alkane_mapping_list}

\end{figure}

\section*{Peptide Sequence Mapping}

\begin{figure*}[!htb]
    \centering
    \includegraphics[width=0.8\linewidth]{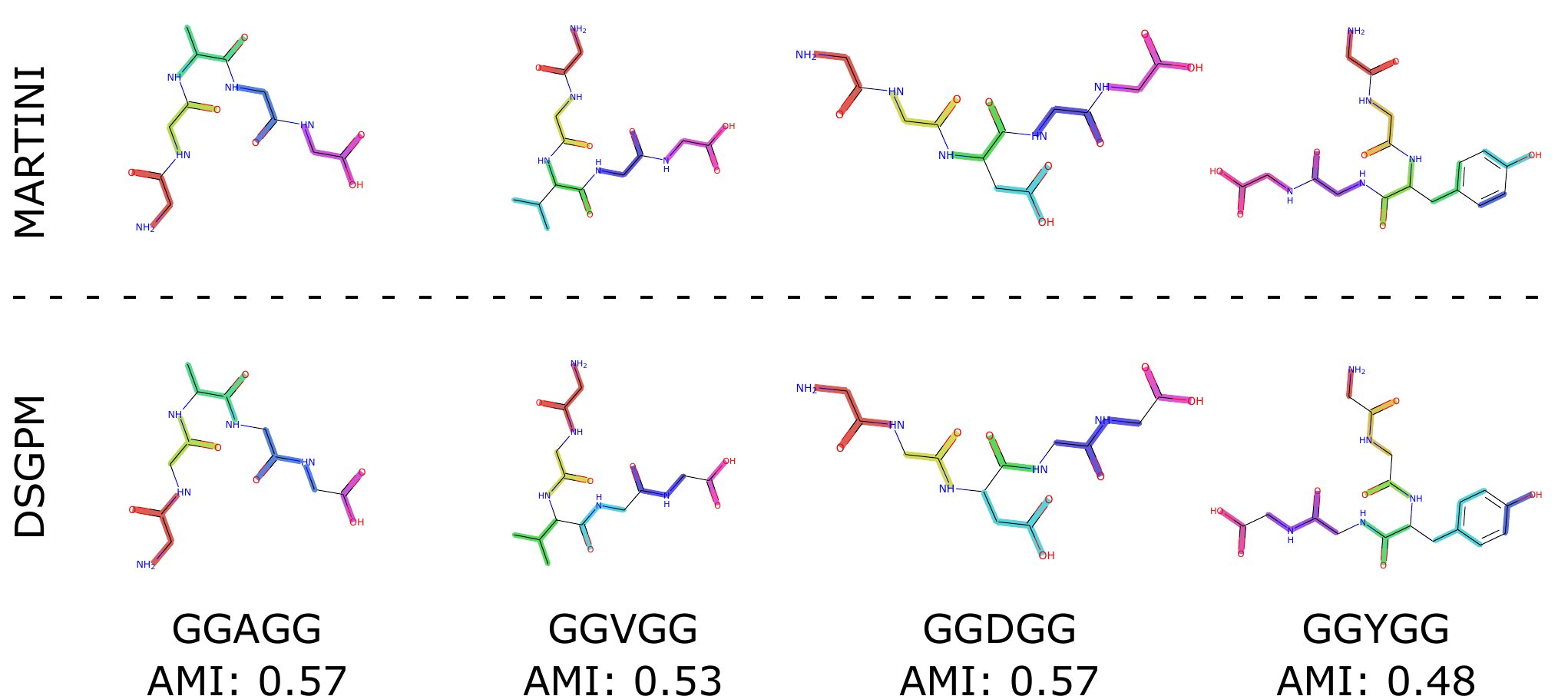}
    \caption{Visualization of CG mappings for 4 peptides. We compare the CG mappings predicted by DSGPM to the corresponding mappings determined by the widely used MARTINI method. Atoms and their adjacent bonds belonging to the same CG bead are highlighted with the same color.}
    \label{fig:martini_comp}
\end{figure*}

We have considered 4 penta-peptides to compare the predicted CG mappings from \MN{} to the corresponding MARTINI CG models. The amino-acid sequence for the 4 peptides is of the form GGXGG, where G is glycine and X is either alanine (A), valine (V), aspartic acid (D) or tyrosine (Y). Note that peptides are previously unseen by the \MN{} model. For each of the peptides, we set the partition number hyperparameter for \MN{} to be equal to the number of CG beads in its MARTINI CG model. The MARTINI CG models for G and A have one bead each, those for V and D have two beads each and the CG model for Y has 4 beads. Hence, the number of partition hyperparameter for \MN{} was set as 5 for GGAGG, 6 for GGVGG and GGDGG, and 8 for GGYGG. Fig. \ref{fig:martini_comp} shows the predicted CG mappings along with the MARTINI CG models for the 4 penta-peptides. The predicted CG mappings closely mirror the MARTINI models for GGAGG, GGVGG and GGDGG, albeit with some deviations. The most prominent difference between the predicted result and the MARTINI CG model is observed for GGYGG.

\clearpage
\small
\bibliographystyle{plainnat}
\bibliography{MLCG}
\end{document}